\documentclass[11pt]{article}

\usepackage[margin=1in]{geometry}
\usepackage{amsmath,amssymb}
\usepackage{booktabs}
\usepackage{array}
\usepackage{tabularx}
\usepackage{enumitem}
\usepackage{xcolor}
\usepackage{hyperref}
\usepackage{microtype}

\hypersetup{hidelinks}
\newcommand{\R}{\mathbb{R}}
\newcommand{\ind}{\mathbf{1}}
\newcommand{\Lhat}{\widehat{\mathcal L}}

\title{Conditional Regime Analog Forecasting with Trajectories:\\
A Nonparametric Framework for Multivariate Probabilistic Time-Series Prediction}
\author{Giancarlo Vercellino\\Independent Researcher}
\date{August 2, 2026}

\begin{document}
\maketitle

\begin{abstract}
We propose Conditional Regime Analog Forecasting with Trajectories (CRAFT), a nonparametric framework for multivariate probabilistic time-series prediction. The method constructs paired backward and forward trajectory profiles from cumulative multi-horizon returns, learns recurrent low-dimensional regime labels in both spaces using singular value decomposition, change-point segmentation, and segment clustering, and estimates a backward-to-forward conditional regime correspondence. Forecast distributions are obtained by sampling historically realized future trajectory profiles according to a composite compatibility score that combines future-regime correspondence with similarity to the current backward trajectory profile. Unlike parametric vector autoregressions or Gaussian state-space models, CRAFT preserves empirical cross-sectional and multi-horizon dependence by resampling complete future paths. We describe the estimator, its diagnostics, and a reproducible simulation benchmark comparing CRAFT with direct analog resampling, SVD analogs, unconditional bootstrap, OLS VAR bootstrap, and random-forest forecasts.
\end{abstract}


\section{Introduction}
Many applied forecasting problems require predictive distributions rather than point forecasts. This is especially true in financial and economic applications, where decision makers care about downside risk, path dependence, cross-asset dependence, and the probability of extreme outcomes. Classical parametric models, including vector autoregressions and Gaussian state-space models, provide useful baselines but often rely on assumptions that are difficult to maintain in environments with nonlinear dependence, changing volatility, and recurring market regimes.

This paper introduces Conditional Regime Analog Forecasting with Trajectories (CRAFT), a nonparametric method for multivariate probabilistic forecasting. CRAFT combines four familiar ideas into a single forecasting architecture. First, each observation is represented by a backward trajectory profile, summarizing recent multi-horizon cumulative returns, and by a forward trajectory profile, recording subsequently realized multi-horizon cumulative returns. Second, both profile spaces are compressed by singular value decomposition (SVD), segmented by change-point detection, and clustered across time to produce recurrent low-dimensional regime labels. Third, the method estimates a backward-to-forward conditional regime correspondence between labels from the backward feature space and labels from the forward feature space. Fourth, the predictive distribution is generated by resampling complete historical forward trajectory profiles, weighted by a composite compatibility score.

The resulting estimator may be viewed as a regime-conditioned analog ensemble. Given the current trajectory profile, CRAFT does not generate artificial paths from an explicit parametric law. Instead, it assigns sampling weights to historically observed future paths and draws from the resulting empirical distribution. This preserves the realized dependence structure across assets and forecast horizons. The approach is therefore particularly natural when realistic joint forecast scenarios are more important than closed-form likelihoods.

\begin{figure}[t]
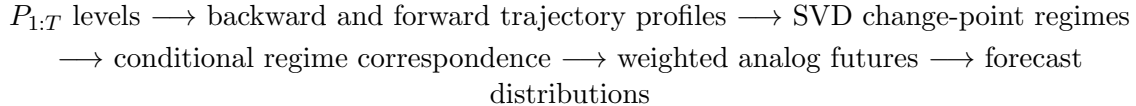

\centering
\fbox{\parbox{0.92\linewidth}{\centering
$P_{1:T}$ levels $\longrightarrow$ backward and forward trajectory profiles $\longrightarrow$ SVD change-point regimes\\[2pt]
$\longrightarrow$ conditional regime correspondence $\longrightarrow$ weighted analog futures $\longrightarrow$ forecast distributions}}
\caption{CRAFT pipeline. Levels are transformed into profile matrices, which are converted into recurrent SVD-based regimes, a conditional correspondence is estimated from backward regimes to forward regimes, and forecast distributions are obtained by reweighting historical future paths.}
\end{figure}

The contribution of CRAFT is not a new change-point algorithm, a new SVD estimator, or a new nearest-neighbor rule. Rather, CRAFT is a dual-partition conditional analog architecture: it partitions the backward and forward profile spaces separately, learns an empirical correspondence from current-profile labels to future-profile labels, and then resamples complete historical future paths. This is a narrower novelty claim than proposing a wholly new forecasting paradigm. More specifically, CRAFT occupies a distinct position relative to three nearby model classes. Unlike Markov-switching models, CRAFT does not assume latent Markov dynamics or parametric emissions. Unlike nearest-neighbor analog forecasting, CRAFT weights analog futures by conditional regime correspondence and backward-profile similarity rather than raw distance alone. Unlike factor forecasting, CRAFT uses SVD factors to define recurrent regimes and analog weights, not to fit a linear predictive equation.

\section{Related Literature}
CRAFT is related to several strands of the time-series and forecasting literature. Regime-switching models, especially the Markov-switching model of Hamilton \cite{hamilton1989}, provide an influential framework for representing discrete changes in economic time-series dynamics. Extensions include regime-switching factor models and multifractal volatility models \cite{kimnelson1998,calvet2006}. CRAFT shares the broad idea that dynamics differ across regimes, but it does not assume a Markov transition matrix over latent states or a parametric observation density. In CRAFT, backward labels $X_t$ and forward labels $Y_t$ are learned in different feature spaces, so the mapping from $X_t$ to $Y_t$ is not a Markov chain over a single state space.

The regime construction step is related to statistical change-point detection. In the implementation studied here, SVD score coordinates are segmented using mean-variance change-point methods, with post-processing to merge short or statistically similar regimes. The PELT algorithm of Killick, Fearnhead, and Eckley \cite{killick2012} is a standard computational reference for efficient change-point detection.

The use of SVD links CRAFT to factor-based forecasting. Principal components are widely used to summarize high-dimensional predictor panels, particularly in macroeconomic forecasting \cite{stockwatson2002}. CRAFT uses SVD not to fit a linear regression directly, but to construct low-dimensional coordinates on which regime segmentation becomes more stable and interpretable.

The sampling step is closest in spirit to analog forecasting and nearest-neighbor prediction. Early analog ideas appear in atmospheric prediction \cite{lorenz1969}, nonlinear time-series prediction \cite{sugihara1990,casdagli1989}, and nonparametric nearest-neighbor forecasting \cite{yakowitz1987}. Analog ensemble forecasting has also been developed extensively in weather forecasting \cite{dellemonache2013}. CRAFT differs from a standard nearest-neighbor analog by conditioning on a learned backward-to-forward regime correspondence and backward-profile similarity rather than directly on Euclidean distance in the original feature space.

Finally, because CRAFT targets full predictive distributions, it is related to conditional distribution and quantile forecasting methods such as quantile regression \cite{koenker1978}, quantile regression forests \cite{meinshausen2006}, and generalized random forests \cite{athey2019}. Predictive distributions may be evaluated using proper scoring rules such as the continuous ranked probability score (CRPS) and other tools discussed by Gneiting and Raftery \cite{gneiting2007}.

\section{Problem Setup}
Let
\[
P_t=(P_{t,1},\ldots,P_{t,m})^\top\in\R^m
\]
denote a multivariate level process observed at times $t=1,\ldots,T$, where $m$ is the number of assets or series. We assume $P_{t,j}\neq0$ when scaled returns are computed. Let $k\geq1$ denote a profile window. For each time $t$, asset $j$, and horizon $h=1,\ldots,k$, define the backward cumulative return
\[
B_{t,j,h}=\frac{P_{t,j}}{P_{t-h,j}}-1
\]
and the forward cumulative return
\[
F_{t,j,h}=\frac{P_{t+h,j}}{P_{t,j}}-1.
\]
Stacking over assets and horizons gives vectors
\[
B_t\in\R^{mk},\qquad F_t\in\R^{mk}.
\]
The backward trajectory profile $B_t$ summarizes recent multi-horizon behavior ending at time $t$, while $F_t$ represents the future path to be forecast from time $t$.

Let $I$ denote the set of valid training origins for which both $B_t$ and $F_t$ are fully observed. The goal is to estimate the conditional predictive law
\[
\mathcal L(F_T\mid B_T),
\]
and, by transformation, the predictive law of future levels
\[
P_{T+h,j}=P_{T,j}(1+F_{T,j,h}),\qquad h=1,\ldots,k.
\]
If a single forecast horizon $H\leq k$ is requested, CRAFT replaces $F_t$ in the target regime construction by the subvector
\[
F_t^{(H)}=(F_{t,1,H},\ldots,F_{t,m,H})^\top.
\]
The final sampler may still retain complete forward trajectory profiles when full path reconstruction is desired.

Let $J_T$ denote the set of times through the forecast origin for which the backward trajectory profile $B_t$ is observable. In rolling use, $R_B$ is fitted on all available backward trajectory profiles $\{B_t:t\in J_T\}$, including the current date, while $R_F$ and the conditional regime correspondence are estimated only on $I$, where future trajectory profiles are known. The canonical CRAFT estimand is the unweighted empirical conditional distribution of realized forward trajectory profiles induced by this fitted partition and correspondence. Volume-weighted and tail-tempered versions are tilted empirical laws and are treated as extensions rather than the default model.

\begin{table}[t]
\centering
\caption{Core CRAFT notation.}
\begin{tabularx}{0.94\linewidth}{>{\raggedright\arraybackslash}p{0.18\linewidth}X}
\toprule
Symbol & Meaning\\
\midrule
$P_t$ & Multivariate level vector observed at time $t$.\\
$B_t$ & Backward trajectory profile ending at time $t$.\\
$F_t$ & Forward trajectory profile realized after time $t$.\\
$X_t$ & Backward-regime label vector $R_B(B_t)$.\\
$Y_t$ & Future-regime label vector $R_F(F_t)$.\\
$X^\ast$ & Current backward-regime label vector $R_B(B_T)$.\\
$\omega_t(B_T)$ & Sampling weight assigned to historical future trajectory profile $F_t$.\\
\bottomrule
\end{tabularx}
\end{table}

\section{The CRAFT Estimator}
\subsection{Trajectory-Profile Construction}
CRAFT begins by constructing the paired empirical sample
\[
\{(B_t,F_t):t\in I\}.
\]
Rows with undefined values, caused for example by insufficient lag or lead observations or by zero scaling denominators, are removed from the correspondence sample. Backward trajectory profiles that are observable through the current date remain available for fitting $R_B$. The current predictor profile is $B_T$, the most recent backward trajectory profile available at the forecast origin.

\subsection{Regime Construction}
CRAFT learns two separate regime maps:
\[
R_B:\R^{mk}\to\mathcal X,\qquad R_F:\R^{d_F}\to\mathcal Y,
\]
where $d_F=mk$ for all-horizon forecasting and $d_F=m$ for a single requested horizon. The backward map $R_B$ is fitted to the matrix with rows $B_t$ for all observable backward trajectory profiles through the forecast date; the forward map $R_F$ is fitted only to rows $F_t$ or $F_t^{(H)}$ for origins with known futures.

For a generic profile matrix $C\in\R^{n\times p}$, the canonical construction is:
\begin{enumerate}[leftmargin=2em]
\item Optionally clean or impute numeric columns, then center and scale each feature:
\[
\widetilde C_{i\ell}=\frac{C_{i\ell}-\mu_\ell}{s_\ell},
\]
with $s_\ell$ bounded below by a small $s_{\min}>0$ when a column has zero or near-zero variance.

\item Compute
\[
\widetilde C=UDV^\top,
\]
select $r$ loading vectors, and form scores
\[
Z=\widetilde C V_r.
\]
The canonical specification uses a fixed $r$; alternative rank rules are extensions.

\item For each selected score coordinate $Z_{\cdot q}$, detect temporal change-points in mean and variance, producing contiguous segments $S_{q,1},\ldots,S_{q,L_q}$.

\item Summarize each segment by stable quantities such as its mean, variance, length, and end-point levels:
\[
h_{q,\ell}=\left(\bar z_{q,\ell},s^2_{q,\ell},|S_{q,\ell}|,z^{\mathrm{first}}_{q,\ell},z^{\mathrm{last}}_{q,\ell}\right).
\]

\item Cluster segment summaries $h_{q,\ell}$ across time. The cluster assignment $g_{q,\ell}\in\{1,\ldots,K_q\}$, rather than the segment index itself, is the reusable regime label. This makes regimes recurrent: similar states occurring in separated historical periods can share the same label.

\item Label each observation by the cluster attached to its containing segment:
\[
R_{i,q}=g_{q,\ell}\quad\text{if }i\in S_{q,\ell},\qquad R(C_i)=(R_{i,1},\ldots,R_{i,r}).
\]
\end{enumerate}

For assigning a new score value $z_q$ to an existing recurrent regime, CRAFT stores each cluster mean $\bar z_{qg}$, scale $\sigma_{qg}$, and empirical segment prior $\pi_{qg}$. The scale is regularized as
\[
\sigma^{\mathrm{reg}}_{qg}=\max(\sigma_{qg},s_{\min}).
\]
A Gaussian-shaped assignment score is
\[
A_{qg}(z_q)=\log\pi_{qg}-\frac12\left(\frac{z_q-\bar z_{qg}}{\sigma^{\mathrm{reg}}_{qg}}\right)^2-\log\sigma^{\mathrm{reg}}_{qg}.
\]
The assigned label is $\arg\max_g A_{qg}(z_q)$. Ties are broken by smaller standardized distance, then by larger prior $\pi_{qg}$, then by the smallest label index. A joint vector $R(C_i)$ that has not appeared historically is still allowed; the conditional correspondence step handles unseen combinations by hierarchical backoff.

A regime is therefore not merely a contiguous time segment and not a cluster in the original feature space. It is a recurrent cluster of time-contiguous SVD factor segments, interpreted as a reusable locally stable state of the trajectory-profile dynamics.

Applying this procedure to backward trajectory profiles gives categorical predictors
\[
X_t=R_B(B_t)=\left(X_t^{(1)},\ldots,X_t^{(Q_B)}\right),
\]
and applying it to forward trajectory profiles gives categorical targets
\[
Y_t=R_F(F_t)=\left(Y_t^{(1)},\ldots,Y_t^{(Q_F)}\right).
\]
In the canonical specification the same fixed rank is used in both profile spaces, so $Q_B=Q_F=r$; extensions may use different ranks. For the forecast origin, the current label vector is
\[
X^\ast=R_B(B_T).
\]

\subsection{Conditional Regime Correspondence}
Because $X_t$ and $Y_t$ are labels from different feature spaces, CRAFT estimates a backward-to-forward regime correspondence rather than a transition matrix for a single Markov chain. For each future-regime coordinate $q=1,\ldots,Q_F$, CRAFT estimates conditional class probabilities over $\mathcal C_q$, the observed classes of $Y_t^{(q)}$.

The canonical model uses unit observation weights. For a subset of backward-regime coordinates $A\subseteq\{1,\ldots,Q_B\}$, define
\[
N^A_{q,c}(X^\ast)=\sum_{t\in I}\ind\{X_{t,A}=X^\ast_A,\;Y_t^{(q)}=c\},
\qquad
n_A(X^\ast)=\sum_{c\in\mathcal C_q}N^A_{q,c}(X^\ast).
\]
The full-key estimator uses $A=\{1,\ldots,Q_B\}$, but exact keys can be sparse. CRAFT therefore uses hierarchical backoff:
\begin{enumerate}[leftmargin=2em]
\item use the full regime vector if its support exceeds $n_{\min}$;
\item otherwise use the largest best-supported subset $A$, breaking ties by support and then by a fixed coordinate order;
\item otherwise use a similar-regime estimator with kernel weights
\[
K_X(X_t,X^\ast)=\exp\left\{-\frac{d_X(X_t,X^\ast)^2}{h_X^2}\right\},
\]
where $d_X$ is a Hamming or validation-weighted Hamming distance on regime vectors;
\item otherwise revert to the unconditional future-regime distribution.
\end{enumerate}

Let $\nu_t(X^\ast)$ denote the nonnegative historical weights produced by the selected backoff level. For exact or subset matching these are indicators; for similar-regime backoff they are kernel weights; for the unconditional level they are all one. With global class shares
\[
\pi^{(0)}_{q,c}=\frac{\sum_{t\in I}\ind\{Y_t^{(q)}=c\}+\alpha}{|I|+\alpha|\mathcal C_q|},
\]
the smoothed correspondence estimate is
\[
\widehat p_q(c\mid X^\ast)=
\frac{\sum_{t\in I}\nu_t(X^\ast)\ind\{Y_t^{(q)}=c\}+\alpha\pi^{(0)}_{q,c}}
{\sum_{t\in I}\nu_t(X^\ast)+\alpha}.
\]
The shrinkage $\alpha$ is controlled by validation or set to a small default. If no backoff level has positive effective support, the estimate is $\pi^{(0)}_{q,c}$.

\subsection{Analog Forecasting Interpretation}
CRAFT is an analog forecasting method because its predictive distribution is a weighted empirical distribution over historically realized future paths. The algorithm does not simulate futures from a parametric law; it reuses observed future trajectory profiles $F_t$, assigning each one a probability according to how compatible its learned future regime is with the current backward regime and how similar its backward trajectory profile is to today's backward trajectory profile. In compact form,
\[
\Lhat(F_T\mid B_T)=\sum_{t\in I}\omega_t(B_T)\delta_{F_t}.
\]
A classical analog estimator assigns weights directly from proximity between historical and current backward trajectory profiles,
\[
\text{classical analog: }\omega_t(B_T)\text{ depends on }d(B_t,B_T).
\]
CRAFT instead uses a composite compatibility score:
\[
\text{CRAFT analog: }\omega_t(B_T)\text{ depends on }
\{\widehat p_q(Y_t^{(q)}\mid X^\ast)\}_{q=1}^{Q_F}
\text{ and }d_B(B_t,B_T).
\]
The aggregation over $q$ is a scoring device, not a joint posterior over the full target vector. A joint posterior interpretation requires an explicit model for $P(Y^{(1)},\ldots,Y^{(Q_F)}\mid X^\ast)$. Thus CRAFT is not merely a search for the closest historical window. It samples historical futures with compatible future-regime labels while retaining analog similarity in the original backward trajectory profile.

\subsection{Composite-Compatible Analog Sampling}
At the forecast origin, the fitted correspondence model produces target-specific probabilities
\[
\widehat p_q(c\mid X^\ast),\qquad c\in\mathcal C_q.
\]
Each historical future trajectory profile $F_t$ is scored according to its realized future-regime vector $Y_t$. Let $a_q\geq0$ be target aggregation weights with $\sum_q a_q=1$. The regime-compatibility component is
\[
r_t=\sum_{q=1}^{Q_F}a_q\log\left\{\widehat p_q\left(Y_t^{(q)}\mid X^\ast\right)\right\}.
\]
To combine regime and distance information, define a standardized backward-profile distance $d_B(B_t,B_T)$ and kernel
\[
K_B(t,T)=\exp\left\{-\frac{d_B(B_t,B_T)^2}{2h_B^2}\right\}.
\]
The canonical composite score is
\[
\eta_t=r_t+\rho\log\{K_B(t,T)+\varepsilon_K\},
\]
where $\rho\geq0$ controls the analog-distance contribution and $\varepsilon_K>0$ avoids $\log0$. Setting $\rho=0$ gives a pure regime-correspondence sampler; larger $\rho$ makes within-regime analog similarity more important.

The canonical model sets the tail penalty to $\lambda=0$. A robustness variant may use
\[
\eta_t^{(\lambda)}=\eta_t-\lambda s_t,
\]
where $s_t$ is a robust extremeness score of $F_t$, but this is not the default because penalizing extreme historical paths can weaken value-at-risk, expected-shortfall, and stress-scenario forecasts.

The final sampling probability assigned to historical future trajectory profile $t$ is
\[
\omega_t(B_T)=\frac{\exp(\eta_t^{(\lambda)})}{\sum_{i\in I}\exp(\eta_i^{(\lambda)})},
\qquad \lambda=0\text{ in the canonical model}.
\]
The CRAFT predictive law is the discrete empirical distribution
\[
\Lhat_{\mathrm{CRAFT}}(F_T\mid B_T)=\sum_{t\in I}\omega_t(B_T)\delta_{F_t},
\]
where $\delta_{F_t}$ denotes a point mass at the observed future trajectory profile $F_t$. Drawing $M$ samples from this distribution yields analog future trajectory profiles
\[
F_T^{(1)},\ldots,F_T^{(M)}\sim\Lhat_{\mathrm{CRAFT}}(F_T\mid B_T).
\]
Because each draw is an entire historical forward trajectory profile, the method preserves empirical dependence across assets and horizons.

\subsection{Forecast Reconstruction}
For each draw $b$, asset $j$, and horizon $h$, the sampled cumulative return is transformed into a level forecast by
\[
P_{T+h,j}^{(b)}=P_{T,j}\left(1+F_{T,j,h}^{(b)}\right).
\]
The empirical samples $\{F_{T,j,h}^{(b)}\}_{b=1}^{M}$ and $\{P_{T+h,j}^{(b)}\}_{b=1}^{M}$ define return and level forecast distributions. Smooth marginal density, distribution, and quantile functions may be obtained by fitting smoothed empirical quantile or density approximations to these samples \cite{silverman1986}.

\section{Algorithm}
\begin{center}
\fbox{\begin{minipage}{0.94\linewidth}
\textbf{Canonical CRAFT procedure}\par\medskip
\textbf{Inputs.} Multivariate levels $P_{1:T}$; profile window $k$; number of draws $M$; fixed SVD dimension $r$; change-point and recurrent segment-clustering settings; correspondence shrinkage $\alpha$; backoff threshold $n_{\min}$; analog-distance weight $\rho$. Canonical defaults are unit observation weights and $\lambda=0$.\par\medskip
\textbf{Outputs.} Predictive return and level distributions; sampled paths; regime labels; correspondence, support, and entropy diagnostics; rolling-backtest results.\par\medskip

\begin{enumerate}[leftmargin=2.2em,itemsep=0.5em]
\item \textbf{Profiles.} Construct backward profiles $B_t$ for all observable origins through $T$, and forward profiles $F_t$ for origins $t\in I$ with known futures.
\item \textbf{Regimes.} Fit $R_B$ to all observable backward profiles through $T$. Fit $R_F$ to the known forward-profile matrix (or a selected-horizon submatrix). Compute $X_t=R_B(B_t)$, $Y_t=R_F(F_t)$, and $X^\ast=R_B(B_T)$.
\item \textbf{Correspondence.} For each target coordinate $q$, estimate $\widehat p_q(c\mid X^\ast)$ using the hierarchy full key $\rightarrow$ subset $\rightarrow$ similar regime $\rightarrow$ unconditional. Set target weights $a_q$, with $a_q=1/Q_F$ by default unless validation supports alternatives.
\item \textbf{Sampling.} For every historical candidate $t\in I$, compute $r_t$, $K_B(t,T)$, and the composite score $\eta_t$; normalize to $\omega_t(B_T)$. Draw $M$ complete future profiles with replacement from $\{F_t:t\in I\}$ using these probabilities.
\item \textbf{Forecast.} Reconstruct return and level forecast distributions from the sampled trajectories and return the distributions, paths, labels, diagnostics, and rolling-backtest results.
\end{enumerate}
\medskip
\[
P_{1:T}\longrightarrow(B_t,F_t)\longrightarrow(X_t,Y_t,X^\ast)
\longrightarrow\widehat p_q(\cdot\mid X^\ast)
\longrightarrow\omega_t(B_T)
\longrightarrow\{F_T^{(b)}\}_{b=1}^{M}.
\]
\end{minipage}}
\end{center}

\section{Computational Complexity}
Let $n=|I|$, $p=mk$, and let $r$ be the number of retained SVD factors in each trajectory profile space. Constructing trajectory profiles costs $O(nmk)$. A truncated SVD costs approximately $O(npr)$ per profile matrix, while change-point detection and recurrent segment clustering across retained scores cost roughly $O(nr)$ plus the clustering cost for segment summaries. Exact correspondence counts cost $O(nQ_F)$. Hierarchical backoff using precomputed subset supports costs $O(nQ_FQ_B)$ for the common nested-subset implementation, with an exhaustive subset search bounded by $O(nQ_F2^{Q_B})$. Prediction at a single origin costs $O(n(Q_F+p))$ to score all historical analogs and $O(M)$ to sample $M$ paths after normalization.

\section{Diagnostics and Model Selection}
CRAFT exposes diagnostics that are useful for model selection, support assessment, and forecast interpretation.

\subsection{Correspondence Accuracy}
The historical sample is split into training and validation segments. For each target regime coordinate, the model reports classification accuracy, balanced accuracy, weighted accuracy, and weighted balanced accuracy for the conditional correspondence. If $\widehat Y_t^{(q)}$ is the modal predicted class for validation observation $t$, ordinary accuracy is
\[
\mathrm{Acc}_q=\frac{1}{|V|}\sum_{t\in V}\ind\{\widehat Y_t^{(q)}=Y_t^{(q)}\}.
\]
Balanced accuracy averages recall over target classes and is useful when future regimes are imbalanced. The implementation also reports joint validation accuracy for the full target label vector, while recognizing that the canonical score is composite rather than a fitted joint posterior.

\subsection{Regime Support Diagnostics}
CRAFT reports support diagnostics for every rolling forecast origin:
\begin{itemize}[leftmargin=2em]
\item number of distinct backward regime keys and future regime labels;
\item observations per full key and per selected backoff subset;
\item percentage of forecasts using exact full-key matches;
\item percentage using partial backoff or similar-regime backoff;
\item percentage reverting to the unconditional distribution;
\item effective number of analog paths, $(\sum_t\omega_t^2)^{-1}$.
\end{itemize}
These diagnostics are central. Sparse regime keys can otherwise make the estimator appear more precise than its support warrants.

\subsection{Correspondence Entropy}
For each target coordinate, the entropy of the fitted correspondence distribution is
\[
H_q(X^\ast)=-\sum_{c\in\mathcal C_q}\widehat p_q(c\mid X^\ast)\log\widehat p_q(c\mid X^\ast).
\]
Large entropy indicates diffuse future-regime correspondence; small entropy indicates concentrated correspondence.

\subsection{Sampler Concentration}
If target weights $a_q$ are derived from validation metrics, CRAFT reports an effective number of weighted targets:
\[
N_{\mathrm{eff}}^{(a)}=\frac{1}{\sum_q\widetilde a_q^2},
\qquad
\widetilde a_q=\frac{a_q}{\sum_r a_r}.
\]
This diagnostic indicates whether the final sampler is driven by many regime coordinates or only by a small subset.

\subsection{Backtesting Metrics}
For rolling-origin evaluation, CRAFT refits the model at selected historical origins and compares predictive draws with realized future outcomes. For each asset and horizon, one may compute median absolute error, root mean squared error, directional accuracy, central interval coverage, PIT-style rank diagnostics, CRPS, and pinball loss. Joint-distribution quality should also be evaluated using the energy score, variogram score, cross-asset correlation error, tail co-exceedance error, and portfolio value-at-risk and expected-shortfall calibration.

If $\widehat q_{\alpha,t,j,h}$ denotes the predicted $\alpha$-quantile of the cumulative return and $r_{t,j,h}$ is the realized return, the pinball loss is
\[
L_\alpha(r,\widehat q)=(\alpha-\ind\{r<\widehat q\})(r-\widehat q).
\]
Coverage of a nominal 90\% interval is estimated by
\[
\widehat C_{0.90}=\frac{1}{N}\sum_i\ind\{\widehat q_{0.05,i}\leq r_i\leq\widehat q_{0.95,i}\}.
\]
Forecast comparisons should report confidence intervals and Diebold--Mariano-style tests \cite{dieboldmariano1995} where appropriate, with standard errors adjusted for overlapping forecast horizons.

\section{Simulation Benchmark and Empirical Reporting}
This section reports a reproducible simulation benchmark using the CRAFT package implementation. The benchmark is intended as a controlled implementation check and a first comparison against ordinary analog and probabilistic baselines. It should not be interpreted as final empirical market evidence: the current benchmark does not include a locked external data set. A full empirical study should additionally evaluate CRAFT on real multivariate forecasting tasks, preferably including equity indices, sector ETFs, rates, foreign exchange, commodities, or macro-financial panels.

\subsection{Simulation Design}
The benchmark is generated by \texttt{CRAFT/inst/benchmarks/compare\_craft\_baselines.R}. The script uses fixed random seeds and writes all result files to \texttt{CRAFT/inst/benchmarks/results/}. The simulated panel has $n=240$ observations and $m=3$ assets. Returns are generated from three recurring regimes with regime-specific mean vectors, volatility vectors, and cross-asset correlation matrices, plus a first-order carryover term equal to 0.18 times the previous return vector. Levels are obtained by compounding $1+r_t$. The data-generating seed is 20260802.

The rolling-origin experiment uses profile window $k=8$, forecast horizon $H=5$, 250 predictive draws per forecast, and 18 rolling origins between observations 120 and 234. At each origin, CRAFT is refitted on the data available up to that origin. The CRAFT specification is \texttt{nfactors = 1}, minimum segment length 8, at most 3 regimes per factor, validation fraction 0.20 with a last-block validation split, unit sampler weights, and tail-penalty parameter $\lambda=0$. The script also runs the package unit tests separately through \texttt{testthat}; the local test suite passes under R 4.5.1.

\subsection{Baselines}
Direct analog resampling in the backward-profile space is the primary baseline, because it is the cleanest test of whether future-regime modelling adds value beyond ordinary analog selection. The implemented benchmark compares CRAFT with five available alternatives:
\begin{itemize}[leftmargin=2em]
\item \textbf{Direct analog:} kernel-weighted resampling of historical future profiles using standardized Euclidean distance between the current backward trajectory profile and historical backward trajectory profiles.
\item \textbf{SVD analog:} the same analog sampler applied after projecting backward trajectory profiles into principal-component score space.
\item \textbf{Unconditional bootstrap:} uniform resampling of historical future profiles.
\item \textbf{OLS VAR bootstrap:} a base-R vector autoregression fitted to log returns, with residual resampling and path simulation to horizon $H$.
\item \textbf{Random forest:} an optional tree ensemble baseline using \texttt{randomForest}, where per-tree predictions provide an empirical draw distribution for each target asset.
\end{itemize}

\begin{table}[t]
\centering
\caption{Simulation benchmark: marginal forecast quality and interval calibration.}
\small
\begin{tabular}{lrrrrr}
\toprule
Model & CRPS & Pinball & Directional & 90\% cover. & 90\% width\\
\midrule
Unconditional bootstrap & 0.02285 & 0.00876 & 0.500 & 0.926 & 0.159\\
Direct analog & 0.02330 & 0.00895 & 0.481 & 0.889 & 0.144\\
CRAFT & 0.02378 & 0.00915 & 0.481 & 0.907 & 0.160\\
SVD analog & 0.02389 & 0.00879 & 0.352 & 0.852 & 0.140\\
OLS VAR bootstrap & 0.02402 & 0.00875 & 0.444 & 0.907 & 0.151\\
Random forest & 0.02529 & 0.00957 & 0.463 & 0.833 & 0.133\\
\bottomrule
\end{tabular}
\end{table}

\begin{table}[t]
\centering
\caption{Simulation benchmark: joint-distribution and tail diagnostics.}
\small
\begin{tabular}{lrrrrr}
\toprule
Model & Energy & Variogram & Corr. err. & VaR breach & Tail Brier\\
\midrule
Unconditional bootstrap & 0.04408 & 0.01425 & 0.156 & 0.111 & 0.10266\\
Direct analog & 0.04614 & 0.01368 & 0.183 & 0.111 & 0.11112\\
CRAFT & 0.04609 & 0.01392 & 0.131 & 0.111 & 0.10559\\
SVD analog & 0.04597 & 0.01367 & 0.155 & 0.167 & 0.10537\\
OLS VAR bootstrap & 0.04630 & 0.01531 & 0.145 & 0.111 & 0.09280\\
Random forest & 0.05044 & 0.02062 & 0.540 & 0.167 & 0.11075\\
\bottomrule
\end{tabular}
\end{table}

\subsection{Simulation Results}
Tables 2 and 3 report the rolling-origin averages. Lower values are better for CRPS, pinball loss, energy score, variogram score, cross-asset correlation error, and tail co-exceedance Brier loss. Higher values are better for directional accuracy, while the nominal target for interval coverage is 0.90.

On this simulation, CRAFT does not dominate the simpler baselines. The unconditional bootstrap has the lowest average CRPS and energy score, while direct analog resampling slightly outperforms CRAFT on CRPS and pinball loss. CRAFT's strongest result is cross-asset dependence: its average correlation error is 0.131, lower than direct analog resampling at 0.183 and lower than the unconditional bootstrap at 0.156. CRAFT also improves tail co-exceedance Brier loss relative to direct analog resampling, 0.10559 versus 0.11112, although the OLS VAR bootstrap has the lowest tail Brier score in this particular run.

\subsection{Support Diagnostics}
Table 4 reports regime-support and sampler diagnostics from the CRAFT forecasts. In all 18 rolling origins, the current backward regime combination had historical support and the unseen-label diagnostic was zero. The number of backward regime keys was small, between 2 and 3, which is expected under the deliberately compact simulation setting.

\subsection{Statistical Comparisons}
Table 5 reports paired loss comparisons of CRAFT against the primary direct-analog baseline. The reported mean improvement is direct-analog loss minus CRAFT loss, so positive values favor CRAFT. Standard errors are heteroskedasticity-and-autocorrelation robust with lag $H-1=4$ to account for overlapping forecast horizons.

None of the CRAFT-versus-direct-analog differences is statistically significant in this small simulation. The random forest baseline is significantly worse than direct analog resampling on CRPS and energy score in the same comparison file, but those results should also be interpreted cautiously because the experiment has only 18 forecast origins.

\begin{table}[t]
\centering
\caption{CRAFT support diagnostics in the simulation benchmark.}
\small
\begin{tabular}{lrrr}
\toprule
Diagnostic & Mean & 5\% & 95\%\\
\midrule
Number of backward regime keys & 2.72 & 2.00 & 3.00\\
Current-key observations & 65.94 & 33.00 & 104.05\\
Unique sampled historical paths & 126.44 & 104.80 & 147.60\\
Sampler probability effective $N$ & 161.00 & 109.95 & 212.05\\
Posterior entropy & 0.193 & 0.000 & 0.724\\
Unseen current-label count & 0.00 & 0.00 & 0.00\\
\bottomrule
\end{tabular}
\end{table}

\begin{table}[t]
\centering
\caption{Paired CRAFT comparisons against direct analog resampling. Positive improvements favor CRAFT.}
\small
\begin{tabular}{lrrrr}
\toprule
Metric & Mean improvement & HAC s.e. & $t$ & $p$-value\\
\midrule
CRPS & -0.000482 & 0.000784 & -0.615 & 0.547\\
Pinball & -0.000199 & 0.000383 & -0.520 & 0.610\\
Energy score & 0.000045 & 0.001135 & 0.040 & 0.969\\
Variogram score & -0.000245 & 0.000903 & -0.271 & 0.790\\
Tail co-exceedance Brier & 0.005524 & 0.005245 & 1.053 & 0.307\\
\bottomrule
\end{tabular}
\end{table}

\subsection{Reproducibility Files}
The benchmark can be reproduced from the repository root with:
\begin{center}
\texttt{Rscript CRAFT/inst/benchmarks/compare\_craft\_baselines.R}
\end{center}
The script writes the following CSV files:
\begin{center}
\begin{tabular}{l}
\texttt{benchmark\_config.csv}\\
\texttt{model\_summary.csv}\\
\texttt{forecast\_metrics\_by\_origin.csv}\\
\texttt{comparison\_tests.csv}\\
\texttt{craft\_regime\_support\_diagnostics.csv}\\
\texttt{availability.csv}
\end{tabular}
\end{center}
The benchmark configuration file records $n=240$, $m=3$, $k=8$, $H=5$, 250 draws, and 18 rolling origins. The availability file records which optional baselines were runnable in the local R environment.

Future empirical tables should retain the same reporting structure but replace the simulation panel with locked real data, expand the number of rolling origins, and add additional probabilistic baselines once their dependencies and tuning protocols are fixed.

\section{Extensions}
The canonical model uses unit observation weights, a fixed SVD rank, recurrent segment-cluster regimes, hierarchical backoff, backward-profile similarity, and no tail penalty. Several implementation features are best treated as optional extensions:
\begin{itemize}[leftmargin=2em]
\item \textbf{Volume weighting.} If volume data $V_t$ are available, correspondence counts or analog scores may be tilted by $w_t\propto(\sum_jV_{t,j}+\varepsilon)^\gamma$. This estimates a volume-weighted empirical law rather than the canonical unweighted law.
\item \textbf{Alternative rank and segmentation rules.} Explained variance, BIC, elbow, Kaiser, regime strength, hybrid rank rules, or alternative segmentation algorithms can replace the fixed-rank canonical choice.
\item \textbf{Tail regularization.} A penalty $\lambda s_t$ may damp extremely unusual historical futures, but $\lambda=0$ is the default because tail penalties can degrade VaR, expected-shortfall, and stress-scenario forecasts.
\item \textbf{Joint target modelling.} A classifier for $P(Y^{(1)},\ldots,Y^{(Q_F)}\mid X^\ast)$ can replace the composite score when sufficient support exists.
\item \textbf{Calibration overlays.} Conformal calibration or other finite-sample interval corrections can be applied to the sampled forecast distribution.
\end{itemize}

\section{Discussion}
CRAFT has several practical advantages. It is nonparametric in the final forecast distribution, it directly produces multivariate scenarios, and it separates the problem of regime identification from the problem of conditional resampling. Because the final predictive law is a weighted empirical distribution over realized future trajectory profiles, dependence across assets and horizons is inherited from the historical sample rather than imposed by a covariance model.

The method also has limitations. First, analog reuse constrains the support of the predictive distribution: CRAFT cannot invent completely unseen future structures, and it can only reweight historical futures already present in the training sample. This makes it robustly empirical but potentially conservative under structural breaks or genuinely novel market states. Second, the quality of the conditional regime correspondence depends on the support of the observed regime keys and the quality of the backoff rule. If the current backward regime is rare or previously unseen, the estimator may rely on subset, similar-regime, or unconditional information. Third, the method involves several tuning parameters, including the profile window, SVD dimension, minimum segment length, recurrent-regime clustering, backoff thresholds, shrinkage, validation split, and analog-distance weight. These parameters should be chosen by rolling validation rather than in-sample fit.

The approach is extensible, but empirical claims should be tied to the canonical specification before optional variants are layered on top. This separation is important because volume weighting, tail regularization, and alternative rank rules change the estimand as well as the implementation.

\section{Conclusion}
This paper presented Conditional Regime Analog Forecasting with Trajectories, a nonparametric algorithm for multivariate probabilistic forecasting. CRAFT constructs backward and forward return trajectory profiles, maps them into recurrent SVD-based regimes, estimates a smoothed backward-to-forward conditional correspondence, and samples complete historical future paths according to composite compatibility. The method provides an interpretable bridge between regime modeling and analog ensemble forecasting, with diagnostics for correspondence uncertainty, regime support, sampler concentration, and rolling forecast performance.

CRAFT should be viewed as a modular dual-partition conditional analog architecture rather than a replacement for all parametric models. Its main strength is the production of realistic joint scenarios under nonstationary and regime-dependent dynamics. Future work should provide large-scale benchmark evidence and formal analysis for adaptive partitions under suitable mixing and recurrence assumptions.

\section*{Code Availability}
The implementation is publicly available as the \texttt{CRAFT} R package on CRAN at
\begin{center}
\url{https://cran.r-project.org/web/packages/CRAFT/index.html}.
\end{center}
The exported user-facing functions include \texttt{craft\_fit()} and \texttt{craft\_predict()}. The package source contains the reproducible simulation benchmark at \texttt{inst/benchmarks/compare\_craft\_baselines.R}; in a source checkout, running this script regenerates every simulation table reported in Tables 2--5 and writes the underlying CSV files to \texttt{inst/benchmarks/results/}. The benchmark uses fixed seeds and records package availability in \texttt{availability.csv}. A future empirical release should add a locked empirical data set, a dependency lockfile or session-info record, and scripts reproducing every figure and table on real data.

\bibliographystyle{plain}
\bibliography{references}

\end{document}